\documentstyle[12pt]{article}
\newcommand{\Frac}[2]%
{{\textstyle \frac{\mbox{\footnotesize $#1$}\rule[-0.9mm]{0mm}{1mm}}%
{\mbox{\footnotesize $#2$}\rule{0mm}{3.1mm}}}}
\def\poe{ 
{\cal P} e^{i \int_{y_-}^{0} dy_{-} A_{+} } }

\renewcommand{\thefootnote}{\fnsymbol{footnote}}

\begin{document}
\begin{titlepage}
\vspace*{-12 mm}
\noindent
\begin{flushright}
\begin{tabular}{l@{}}
TUM/T39-01-01 \\
hep-ph/0102036 \\
\end{tabular}
\end{flushright}
\vskip 12 mm
\begin{center}
{\Large \bf Spin structure in non-forward partons } \\
\vspace{3ex}
{\bf Steven D. Bass}
\footnote[2]{steven.bass@cern.ch}
\\[10mm]   
{\em 
Physik Department, 
Technische Universit\"at M\"unchen, \\
D-85747 Garching, Germany }
\\[5mm]

\end{center}
\vskip 10 mm
\begin{abstract}
\noindent
Renormalisation induces anomalous contributions in light-cone correlation 
functions. We discuss the role of the axial anomaly and gluon topology in 
non-forward parton distributions noting that non-forward matrix elements 
of the gluonic Chern-Simons current $K_{\mu}$ are not gauge-invariant even 
in perturbation theory. 
The axial anomaly means that one has to be careful how to interpret 
information from hard exclusive reactions about the orbital angular-momentum 
carried by the proton's internal constituents.

\end{abstract}
\end{titlepage}
\renewcommand{\labelenumi}{(\alph{enumi})}
\renewcommand{\labelenumii}{(\roman{enumii})}
\renewcommand{\thefootnote}{\arabic{footnote}} 
\newpage
%

\section{Introduction}

Non-forward parton distributions \cite{radprd,jidvcs,radrev,jirev,sjb} 
have recently been proposed as tools to study hard exclusive 
reactions:  deeply virtual Compton scattering and exclusive 
meson production \cite{guichon,piller} in $\gamma^* p$ reactions at 
large $Q^2$ (the incident photon virtuality), 
large $s$ (centre of mass energy) 
and small $t$ (squared momentum transfer).
It has further been proposed that deeply virtual Compton scattering 
(DVCS) could provide information about quark and gluonic orbital 
angular-momentum in the proton \cite{jidvcs}.
It is now well established that the spin structure of the proton is 
sensitive to the physics of the axial anomaly and dynamical $U_A(1)$
symmetry breaking in QCD -- for reviews see \cite{epja,reyarev,shorerev}.
In this paper we discuss the role of these phenomena in non-forward
parton distributions and in deeply virtual Compton scattering.
The important details are the gauge dependence of non-forward matrix
elements of the anomalous Chern Simons current $K_+$ -- even in 
perturbation theory -- and the role of zero modes in dynamical $U_A(1)$ 
symmetry breaking.

Perturbative QCD factorisation applies to DVCS.
In perturbative QCD the amplitudes for deeply virtual Compton 
scattering and hard exclusive meson production at small $t$ 
may be written as the integral over the product of non-forward parton 
distributions and coefficient functions [1-5,11].
The non-forward parton distributions describe the momentum and spin
distribution of partons in the target proton and the coefficient functions 
describe the infra-red safe hard $\gamma^*$--parton interaction, which may 
be calculated in perturbative QCD.
Working in light-cone gauge $A_+=0$ the spin independent and spin dependent 
non-forward parton distributions are written as the Fourier transform of 
light-cone correlation functions: the non-forward matrix elements of 
``${\bar q}(0) \gamma_+ q(y)$'' and ``${\bar q}(0) \gamma_+ \gamma_5 q(y)$'' 
respectively at $y_+=y_{\perp}=0$.

The unpolarised DVCS cross-section receives contributions from
both the spin-dependent and spin-independent parts of the DVCS
amplitude ${\cal A}_{\rm DVCS}$ \cite{jidvcs}.
Experiments with polarised beams and targets may help to resolve
the separate spin-dependent and spin-independent contributions
to ${\cal A}_{\rm DVCS}$ \cite{diehl,belitsky}. 
Here we assume that both contributions can be extracted from 
future experimental data and investigate the role of the axial 
anomaly in the interpretation of the data.

In QCD the bare light-cone correlation functions are ultra-violet 
divergent requiring renormalisation \cite{llewellyn}. 
This renormalisation induces the anomalous dimensions 
($Q^2$ dependence of the distributions). 
It also means that the spin dependent 
(non-forward) parton distribution is 
sensitive to the axial anomaly \cite{adler,bell,schwinger}. 
It is important to check whether any non-perturbative physics induced by 
the anomaly is consistent with the original perturbative QCD factorisation.

The main results of this paper are the following.
First, the flavour-singlet part of ${\cal A}_{\rm DVCS}^{\rm spin}$ is 
sensitive to the QCD axial anomaly and does not have a simple canonical
interpretation. 
In full QCD (beyond perturbation theory) whether factorisation provides
a complete description of {\it this contribution} is sensitive to the 
dynamics of axial U(1) symmetry breaking -- the famous $U_A(1)$ problem 
of the $\eta'$ mass --
which has the potential to induce zero-mode contributions to the
flavour-singlet axial-charge $g_A^{(0)}$ and its non-forward generalisation.
Second, Ji \cite{jidvcs} has shown that
information about quark and gluon total angular-momentum 
in the proton ($J_z^q$ and $J_z^g$) may be extracted from 
the forward limit of the spin-independent part of ${\cal A}_{\rm DVCS}$.
Through the axial anomaly in QCD, some fraction of the intrinsic spin 
content of the nucleon and of the constituent quark is carried by its 
quark and gluon partons and some fraction is carried by gluon topology 
through a zero-mode \cite{epja}. 
The zero-mode term ${\cal C}$ is a model independent physical
quantity which can, in principle, be measured through $\nu p$
elastic scattering \cite{bass98}.
The axial anomaly and zero-modes 
contribute separately to $S_z^q$ and $L_z^q$ but cancels in the sum.
Possible zero-modes and scheme dependence mean that values of $L_z$ 
extracted from future DVCS data have to be labelled with respect to 
the degrees of freedom and the perturbative QCD scheme dependence used 
to define $S_z^q$.

The plan of the paper is as follows. To motivate our discussion of 
spin in non-forward parton distributions we begin in Section 2 with 
a brief overview of our present understanding of the proton spin 
problem from polarised deep inelastic scattering.
In Section 3 we give a brief overview of non-forward 
parton distributions -- for excellent reviews see \cite{radrev,jirev}.
Section 4 discusses the axial anomaly and Section 5 its role in the 
spin dependent part of ${\cal A}_{\rm DVCS}$. 
In Section 6 we discuss the role of the anomaly and gluon topology 
in the interpretation of information about $L_z$ from DVCS. Finally, 
in Section 7 we make our conclusions.

\section{The spin structure of the proton}

To motivate our discussion of spin effects in DVCS we first briefly 
review what is known about the quark and gluonic intrinsic spin 
structure of the proton from the interpretation of polarised deep 
inelastic scattering data.
We start with the simple sum-rule for the spin $+{1 \over 2}$ 
proton 
\begin{equation}
{1 \over 2} \ =
\ {1 \over 2} \Sigma + L_z + S_z^g .
\end{equation}
Here,
${1 \over 2}\Sigma$ and $S_z^g$ are the quark and gluonic intrinsic spin 
contributions to the nucleon's spin and $L_z$ is the orbital contribution.
One would like to understand the spin decomposition, Eq.(1),
both in terms of the fundamental QCD quarks and gluons and
also in terms of the constituent quark quasi-particles of low-energy QCD.
In relativistic constituent quark models $\Sigma$ is given by 
the flavour-singlet axial-charge $g_A^{(0)}$.
The value of $g_A^{(0)}$ extracted from polarised deep inelastic
scattering experiments is
$g_A^{(0)}|_{\rm pDIS} = 0.2$ -- 0.35 \cite{windmolders}, 
roughly half of the value predicted by the constituent quark models.
In QCD the interpretation of the individual quantities in Eq.(1) is
quite subtle because of the axial anomaly \cite{shorerev}, issues of 
gauge invariance \cite{jaffem,jaffe} and dynamical $U_A(1)$ symmetry 
breaking 
\cite{epja,fritzsch,forte,shorerev}.
In QCD the axial anomaly induces various gluonic contributions 
to $g_A^{(0)}$.
Working in light-cone gauge $A_+=0$ one finds
\cite{ar,et,ccm,bint,epja}
\begin{equation}
g_A^{(0)}
=
\Biggl(
\sum_q \Delta q - 3 {\alpha_s \over 2 \pi} \Delta g \Biggr)_{\rm partons}
+ \ {\cal C}
\end{equation}
Here ${1 \over 2} \Delta q$ and $\Delta g$ are the amount of spin carried
by quark and gluon partons in the polarised proton and ${\cal C}$ measures
the gluon-topological contribution to $g_A^{(0)}$ \cite{bass98}.

In Eq.(2) $\Delta q_{\rm partons}$ is associated with the hard photon 
scattering on quark and antiquarks with low transverse momentum squared,
$k_T^2$ of the order of typical gluon virtualities in the proton, and 
$\Delta g_{\rm partons}$ is associated with the hard photon scattering on 
quarks and antiquarks carrying $k_T^2 \sim Q^2$.
Jet studies and semi-inclusive measurements of the nucleon's 
spin-dependent $g_1$ structure function may be used 
to determine $\Delta g$ and $\Delta q$ for each flavour \cite{reyarev}.
The topology term ${\cal C}$ is associated with dynamical axial U(1) 
symmetry breaking and has support only at Bjorken $x=0$.
It is missed by polarised deep inelastic scattering experiments but could, 
in principle, be measured in elastic $\nu p$ scattering \cite{bass98}.
An example how to obtain a finite value for ${\cal C}$ 
is provided by
Crewther's theory of quark-instanton interactions \cite{crewther}.
There, any instanton induced suppression of $g_A^{(0)}|_{\rm pDIS}$ 
(the axial charge carried by partons with finite momentum fraction
 $x>0$)
is compensated by a shift of axial charge to the zero-mode so that
the total axial-charge $g_A^{(0)}$ including ${\cal C}$ is conserved.

The quark and gluonic parton decomposition of $g_A^{(0)}$ in Eq.(2) 
is factorisation scheme dependent.
Eq.(2) describes the singlet axial charge in the $k_T^2$ 
cut-off parton model and in the AB and JET schemes \cite{ab,jet}.
In the ${\overline {\rm MS}}$ 
scheme the polarised gluon contribution is absorbed into
$\Delta q$ so that
$\Delta q_{\overline {\rm MS}} = 
\Biggl( \Delta q - {\alpha_s \over 2 \pi} \Delta g \Biggr)_{\rm partons}$
\cite{bodwin}.

The QCD $g_A^{(0)}$ is measured by the proton forward matrix element of 
the flavour singlet axial-vector current.
\begin{equation}
2m s_\mu g_A^{(0)} = 
\langle P,S|
\ J^{GI}_{\mu5} \ |P,S \rangle _c
\end{equation}
where $|P,S \rangle$ denotes a proton state with momentum $P$ and spin $S$ and
\begin{equation}
J^{GI}_{\mu5} = \left[ \bar{u}\gamma_\mu\gamma_5u
                  + \bar{d}\gamma_\mu\gamma_5d
                  + \bar{s}\gamma_\mu\gamma_5s \right]_{GI}
\end{equation} 
is the
gauge-invariantly renormalised singlet axial-vector operator.
In each of the ``partons'', AB and JET schemes $\Delta q$ and 
$\Delta g$ are measured by the forward matrix elements of the 
partially conserved axial-vector current and gluonic Chern-Simons 
current contributions to $J_{+5}^{GI}$ 
between parton states in the light-cone gauge $A_+=0$
-- see Sections 4 and 5 below.
In light-cone gauge the forward matrix elements of the plus 
components of these gauge dependent currents are invariant 
under residual gauge degrees of freedom.
However, the non-forward matrix elements of these currents are 
not invariant, even in perturbative QCD.
This means that some of the schemes commonly used to analyse 
polarised deep inelastic data are not gauge invariant when 
``skewed'' or extrapolated away from the forward direction.

In summary,
the axial anomaly brings in gauge invariance issues and the
possibility of zero-momentum contributions to the individual
spin contributions in Eq.(1).
How does this physics enter the theory of deeply virtual Compton
scattering ?

\section{Non-forward parton distributions}

Non-forward parton distributions are defined as the Fourier transforms of 
light-cone correlation functions.
Consider an incident proton with mass $M$ and momentum $P_{\mu}$ which 
gets given momentum $\Delta_{\mu} = (P'-P)_{\mu}$ and emerges into the 
final state with momentum $P'_{\mu}$. 
We use ${\bar P}_{\mu} = {1 \over 2}(P'+P)_{\mu}$ to denote the average 
nucleon momentum. In this paper we follow the notation of Radyushkin 
\cite{radprd}.
We let $x$ denote the fraction of light-cone momentum $k_+ = x P_+$ 
carried by a parton in the incident proton and $\zeta = \Delta_+ / P_+$ 
denote the amount of light-cone momentum transfered to the proton.

In perturbative QCD the spin averaged cross-section for deeply virtual 
Compton scattering receives contributions from both spin independent
and spin dependent non-forward parton distributions.
The formula suggested by perturbative QCD for the deeply 
virtual Compton amplitude $M^{IJ} = {\cal A}_{\rm DVCS}$ 
is: 
\begin{eqnarray}
& & M^{IJ}( {\vec q}_{\perp}, {\vec \Delta}_{\perp}, \zeta )
= \\ \nonumber
& & 
- {\rm e_q}^2 {1 \over 2 {\bar P}^+}  \ \int_{\zeta -1}^{+1} \ dx \ 
\Biggl[
F_q ( x, \zeta, t, \mu^2) C_q (x, \zeta,Q^2,\mu^2)
+ F_g ( x, \zeta, t, \mu^2) C_g (x, \zeta,Q^2,\mu^2)
\\ \nonumber
& & 
\ \ \ \ \ \ \ \ \ \ \ \ \ \ \ \ \ \ \ \ +
{\tilde F}_q ( x, \zeta, t, \mu^2) {\tilde C}_q (x,\zeta,Q^2,\mu^2)
+ {\tilde F}_g (x, \zeta, t, \mu^2) {\tilde C}_g (x, \zeta,Q^2,\mu^2)
\Biggr] \ + \ {\cal O} \biggl( {1 \over Q} \biggr)
\end{eqnarray}
Here $(I,J)$ refer to the polarisation vectors 
($\uparrow$ or $\downarrow$) of the initial and 
final state photons -- here both taken as ${\uparrow}$;
$F_q$ and $F_g$ are the spin-independent non-forward 
quark and gluonic parton distributions and
${\tilde F}_q$ and ${\tilde F}_g$ are the spin-dependent parton 
distributions. 
These distributions are defined in Eqs.(7-10) below; $\mu^2$ 
denotes the renormalisation scale.
In Eq.(5)
\begin{eqnarray}
C_q^{\uparrow \uparrow} & = &  C_q^{\downarrow \downarrow} =
\Biggl( {1 \over x - i \epsilon} + {1 \over x - \zeta + i \epsilon}
\Biggr)
+ O (\alpha_s)
\\ \nonumber
C_q^{\uparrow \uparrow} & = & - C_q^{\downarrow \downarrow} =
\Biggl( {1 \over x - i \epsilon} - {1 \over x - \zeta + i \epsilon}
\Biggr)
+ O (\alpha_s)
\end{eqnarray}
are the non-forward quark coefficient functions with
$C_q^{\uparrow \downarrow} = C_q^{\downarrow \uparrow} =
 {\tilde C}_q^{\uparrow \downarrow} = {\tilde C}_q^{\downarrow \uparrow} = 0$.
The gluonic coefficients start at order $\alpha_s$, 
viz. $C_g, {\tilde C}_g \sim O(\alpha_s)$.
In DVCS $\zeta$ plays the role of the 
Bjorken variable $\zeta = x_{\rm Bj}$. 
Radyushkin \cite{radplb} has derived conditions on the non-forward
distributions which must be satisfied in order that 
QCD factorisation holds beyond perturbation theory.  
One of these conditions is that 
the non-forward parton distributions do not contain any 
singular behaviour at $x = 0$ or $x = \zeta$.
We shall return to this point in Section 5 below.

The spin independent non-forward quark and gluon distributions are:
\begin{eqnarray}
F_q(x,\zeta,t) 
&=& 
\int {d y_- \over 8 \pi} e^{i x P_+ y_- /2}
\langle P' | {\bar q} (0) \gamma_+ \ \poe \ q (y) | P \rangle
\bigg|_{y_+ = y_{\perp} =0}
\\ \nonumber
&=& 
{1 \over 2 {\bar P}_+}
{\bar U}(P')
\Biggl[
H_q (x,\zeta,t) \ \gamma_+ \
+
E_q (x,\zeta,t) \ 
{i \sigma_{+ \alpha} (- \Delta^{\alpha}) \over 2M} \Biggr] \ U(P)
\end{eqnarray}
and
\begin{eqnarray}
F_g(x,\zeta,t) 
&=& 
{1 \over x P_+}  \int {d y_- \over 8 \pi} e^{i x P_+ y_- /2}
\langle P' | G_{+ \alpha} (0) \ \poe \ G^{\alpha}_{\ +} (y) | P \rangle
\bigg|_{y_+ = y_{\perp} =0}
\\ \nonumber
&=& 
{1 \over 2 {\bar P}_+^2}
{\bar U}(P')
\Biggl[ H_g (x,\zeta,t) \ \gamma_+ \
+
E_g (x,\zeta,t) \ 
{i \sigma_{+ \alpha} (- \Delta^{\alpha}) \over 2M} \Biggr] \ U(P)
\end{eqnarray}
respectively.
In the forward limit $\zeta =0$ one recovers the spin independent
quark and gluon distributions measured in deep inelastic scattering:
$(q \pm {\bar q})(x) = {1 \over 2} ( H_q(x,0,0) \mp H_q (-x,0,0) )$
and $g(x) = {1 \over 2} ( H_g(x,0,0) - H_g (-x,0,0) )$ respectively.
The spin dependent distributions are:
\begin{eqnarray}
{\tilde F}_q(x,\zeta,t) 
&=& 
\int {d y_- \over 8 \pi} e^{i x P_+ y_- /2}
\langle P' | {\bar q} (0) \gamma_+ \ \gamma_5 \poe \ q (y) | P \rangle
\bigg|_{y_+ = y_{\perp} =0}
\\ \nonumber
&=& 
{1 \over 2 {\bar P}_+}
{\bar U}(P')
\Biggl[
{\tilde H}_q (x,\zeta,t) \ \gamma_+ \gamma_5 \
+
{\tilde E}_q (x,\zeta,t) \ 
{1 \over 2M} \gamma_5 (- \Delta_{+}) \Biggr] \ U(P)
\end{eqnarray}
and
\begin{eqnarray}
{\tilde F}_g(x,\zeta,t) 
&=& 
-{i \over x P_+} \int {d y_- \over 8 \pi} e^{i x P_+ y_- /2}
\langle P' | 
G_{+ \alpha} (0) \ \poe \ {\tilde G}^{\alpha}_{\ +}(y_-) | P \rangle
\bigg|_{y_+ = y_{\perp} =0}
\\ \nonumber
&=& 
{1 \over 2 {\bar P}_+^2}
{\bar U}(P')
\Biggl[
{\tilde H}_g (x,\zeta,t) \ \gamma_+ \gamma_5 \
+
{\tilde E}_g (x,\zeta,t) \ 
{1 \over 2M} \gamma_5 (- \Delta_+) \Biggr] \ U(P)
\end{eqnarray}
In the forward limit $\zeta =0$ one recovers the spin dependent quark and 
gluon distributions measured in deep inelastic scattering:
$\Delta (q \pm {\bar q})(x) 
= {1 \over 2} ( {\tilde H}_q(x,0,0) \pm {\tilde H}_q (-x,0,0) )$ and 
$\Delta g(x) = {1 \over 2} ( {\tilde H}_g(x,0,0) + {\tilde H}_g (-x,0,0) )$ 
respectively.
The isotriplet combination $({\tilde E}_u - {\tilde E}_d)$ 
contains the pion pole and the flavour-singlet combination 
$({\tilde E}_u + {\tilde E}_d + {\tilde E}_s)$ 
is sensitive to the axial U(1) problem \cite{crewther,thooft}
through the $\eta'$ pole in the pseudoscalar form-factor.

We now focus on the spin-dependent non-forward quark distribution 
${\tilde F}_q$ to discuss the construction and renormalisation of 
these distributions.
The parton model and the light-cone correlation functions are normally 
formulated in the light-cone gauge $A_+=0$. 
Here the path-ordered exponential becomes a trivial unity factor and is 
dropped from the formalism:
\begin{equation}
\langle P' | {\bar q} (0) \ \gamma_+ \ \gamma_5 \ \poe \ q (y) | P \rangle
\ \ \mapsto \ \
\langle P' | {\bar q} (0) \ \gamma_+ \ \gamma_5 \ q (y) | P \rangle
\end{equation}
In semi-classical QCD, before we come to discuss renormalisation and 
anomalies, the gluonic degrees of freedom have dropped out along with
the path-ordered exponential -- hence the terminology ``quark distribution''.
Before we consider subtleties associated with anomaly theory -- see Section 4
below --
the non-forward parton distributions have the following simple 
interpretation \cite{jirev,sjb}.
Expanding out the quark and gluon field operators in $q(y)$ and 
$G_{\mu \nu}(y)$ one finds the following.
For $\zeta < x <1$ we have the situation where one removes a quark 
carrying light-cone momentum 
$k_+ = x P_+$ and transverse momentum ${\vec k}_{\perp}$
from the initial state proton and re-inserts it into the 
final state proton with the same chirality but with light-cone 
momentum fraction $x- \zeta$ and transverse momentum 
${\vec k}_{\perp} - {\vec \Delta}_{\perp}$.
For $\zeta -1 < x < 0$ one finds the situation for removing an
antiquark with momentum fraction $\zeta - x$ and re-inserting it
with momentum fraction $-x$.
For $0 < x< \zeta$ the photons scatter off a virtual 
quark-antiquark pair in the initial proton wavefunction:
the quark of the pair has light-cone momentum fraction $x$ and 
transverse momentum ${\vec k}_{\perp}$, whereas the antiquark
has light-cone momentum fraction $\zeta - x$ and 
transverse momentum ${\vec \Delta}_{\perp} - {\vec k}_{\perp}$.
The third region $0 < x < \zeta$ is not present in deep inelastic 
scattering where $\zeta=0$.
The points $x=0$ and $x=\zeta$ correspond to zero-momentum modes.
The flavour-singlet spin-dependent parton distributions at these 
two points are sensitive to the role of zero-modes in dynamical
$U_A(1)$ symmetry breaking \cite{crewther,thooft},
which have the potential to generate new non-perturbative 
contributions to the spin-dependent part of ${\cal A}_{DVCS}$ 
beyond those appearing in the factorisation formula (5).
The $J=0$ Regge fixed pole \cite{fixedpoles} 
in the spin-independent part of ${\cal A}_{DVCS}$ 
which generates a constant real term in ${\cal A}_{\rm DVCS}$ is
manifest in Eqs.(5,6) through the $\zeta \rightarrow 0$ limit of $C_q$.

The $x$-moments, $\int_{\zeta -1}^{+1} dx \ x^n$, of the non-forward 
parton distributions are evaluated as follows. 
First one writes $x^{n}$ as a derivative (in $y_-$) acting on 
$e^{i x P_+ y_- / 2}$.
Integrating by parts (with respect to $x$) over the exponential 
yields a Dirac delta-function $\delta(y_-)$. 
The $y_-$ integral then projects out the non-forward matrix elements of 
local operators.
One finds
\begin{eqnarray}
& & \int_{\zeta -1}^1 dx \ x^{n} \ {\tilde F}_q (x,\zeta,t) \\ \nonumber
& & \ \ \ \ \ \ =
{1 \over 2} \biggl( {2 \over P_+} \biggr)^n
\langle P' | {\bar q} (0) \gamma_+ \ \gamma_5 \ (i \partial_+)^n \
q (0) | P \rangle \bigg|_{y_+ = y_{\perp} =0}
\\ \nonumber
& & \ \ \ \ \ \ = 
{1 \over 2 {\bar P}_+}
{\bar U}(P')
\Biggl[
\int_{\zeta -1}^{+1} dx \ x^n \
{\tilde H}_q (x,\zeta,t) \ \gamma_+ \gamma_5 \
+
\int_{\zeta -1}^{+1} dx \ x^n \
{\tilde E}_q (x,\zeta,t) \ 
{1 \over 2M} \gamma_5 (- \Delta_{+}) \Biggr] \ U(P)
\end{eqnarray}
A caveat is due here:
renormalisation means that we have to be careful not to simply set $y_- =0$
in the point-split operator to obtain the local operator. 
In QCD the bare light-cone correlation functions are ultra-violet divergent 
requiring renormalisation \cite{llewellyn}. 
This renormalisation means that spin independent (non-forward) 
parton distribution ${\tilde F}_q$ is sensitive to the axial anomaly.

We now discuss the effect of the anomaly in non-forward parton distributions.

\section{Renormalisation and anomalies}

Going beyond the semi-classical approximation to QCD we have to take into
account that
the renormalised composite operator
\begin{equation}
\biggl[ \ {\bar q} \gamma_{\mu} \gamma_5 q \ \biggr]^{R}_{\mu^2}(0)
\
\neq
\
{\bar q} (0) \gamma_{\mu} \gamma_5 \ {\rm \ multiplied \ by \ } \ q(0)
\end{equation}
This point is especially important when we evaluate the integral 
over $y_-$ of the point-split matrix element with $\delta (y_-)$.
Evaluating the moments of ${\tilde F}_q$ is non-trivial because 
the point-split operator is highly singular in the limit that the 
point splitting is taken to zero.
In full QCD it is necessary to work with renormalised composite operators
instead of their semi-classical prototypes. 
To see this explicitly consider Schwinger's derivation 
\cite{schwinger,jackiwrev} of 
the axial anomaly using point split regularisation.
For $(z'-z'') \rightarrow 0$, one finds that the vacuum to vacuum 
matrix element of the point-split operator in an external gluon field is:
\begin{equation}
\langle {\rm vac} | \
\biggl[ \overline{q}(z') \gamma_{+} \gamma_5 q(z'') \biggr] 
\ | {\rm vac} \rangle
\simeq
{i g \over 8 \pi^2} \
{\tilde G}_{+ \nu} \ {(z'-z'')^{\nu} \over (z'-z'')^2}
\end{equation}
where the gluonic term arises from pinching a gluonic insertion 
between $z'$ and $z''$.
Going to the light-cone ($z_T =0$, $z_+ \rightarrow 0$) the factor
\begin{equation}
{(z'-z'')_+  \over (z'-z'')^2} \ \ \mapsto \ \ {1 \over (z' - z'')_-}
\end{equation}
which diverges when $z' \rightarrow z''$.
One clearly has to be careful and not ensure that the theory and its 
interpretation does not assume equality of both sides in Eq.(13).

The problem is resolved \cite{llewellyn} if, working in light-cone gauge, 
we define
\begin{equation} 
\langle P',S' | 
\biggl[
\overline{q}(0) \gamma_{+} \gamma_{5} q(y_-) \biggr]^R_{\mu^2} 
|P,S \rangle_c
\equiv
\sum_{n} {(i y_- )^n \over n!} 
\langle P',S' | 
\biggl[
\overline{q} \gamma_{+} \gamma_{5} (iD_{+})^{n} q \biggr]^R_{\mu^2} 
(0) |P,S \rangle_c
\end{equation}
and
\begin{equation}
\langle P',S' | 
{\rm Tr} \biggl[ 
G_{+ \nu}(0) {\tilde G}_{\ +}^{\nu}(y_-) 
\biggr]^R_{\mu^2} 
|P,S \rangle_c 
\equiv 
\sum_{n} {(i y_- )^n \over n!}
\langle P',S'| 
{\rm Tr} \biggl[ 
G_{+ \nu} (iD_{+})^{n} {\tilde G}_{+}^{\nu} \biggr]^R_{\mu^2} 
(0) |P,S \rangle_c
\end{equation}
That is, we treat the non-local operators in Eqs.(7-10) as a series 
expansion in terms of renormalised composite local operators in the 
operator product expansion.
The superscript $R$ denotes the renormalisation prescription and 
$\mu^2$ denotes the renormalisation scale. The composite operators
$\biggl[
 \overline{q} \gamma_{+} \gamma_{5} (iD_{+})^{2n} q \biggr](0)$
and
${\rm Tr} \biggl[ 
 G_{+ \nu} (iD_{+})^{2n} {\tilde G}_{+}^{\nu} \biggr](0)$
mix under renormalisation.
The axial vector and higher-spin axial tensor operators are sensitive to the 
axial anomaly.

\subsection{The axial anomaly}

The gauge-invariantly renormalised flavour singlet axial-vector current
\begin{equation}
J^{GI}_{\mu5} = \biggl[ \bar{u}\gamma_\mu\gamma_5u
                  + \bar{d}\gamma_\mu\gamma_5d
                  + \bar{s}\gamma_\mu\gamma_5s \biggr]^{GI}_{\mu^2}
\end{equation} 
satisfies the anomalous divergence equation \cite{adler,bell}
\begin{equation}
\partial^\mu J^{GI}_{\mu5}
= 2f\partial^\mu K_\mu + \sum_{i=1}^{f} 2im_i \bar{q}_i\gamma_5 q_i
\end{equation}
Here
\begin{equation}
K_{\mu} = {g^2 \over 16 \pi^2}
\epsilon_{\mu \nu \rho \sigma}
\biggl[ A^{\nu}_a \biggl( \partial^{\rho} A^{\sigma}_a 
- {1 \over 3} g 
f_{abc} A^{\rho}_b A^{\sigma}_c \biggr) \biggr]
\end{equation}
is a renormalised version of the gluonic Chern-Simons
current and the number of light flavours $f$ is $3$.
Eq.(20) allows us to write
\begin{equation}
J_{\mu 5}^{GI} = J_{\mu 5}^{\rm con} + 2f K_{\mu}
\end{equation}
where $J_{\mu 5}^{\rm con}$ and $K_{\mu}$ satisfy the
divergence equations
\begin{equation}
\partial^\mu J^{\rm con}_{\mu5} 
= \sum_{i=1}^{f} 2im_i \bar{q}_i\gamma_5 q_i
\end{equation}
and
\begin{equation}
\partial^{\mu} K_{\mu} 
= {g^2 \over 8 \pi^2} G_{\mu \nu} {\tilde G}^{\mu \nu}.
\end{equation}
Here
${g^2 \over 8 \pi^2} G_{\mu \nu} {\tilde G}^{\mu \nu}$
is the topological charge density.
The partially conserved current is scale invariant 
and 
the scale dependence of $J_{\mu 5}^{GI}$ is carried entirely
by $K_{\mu}$.
When we make a gauge transformation $U$ 
the gluon field transforms as
\begin{equation}
A_{\mu} \rightarrow U A_{\mu} U^{-1} + {i \over g} (\partial_{\mu} U) U^{-1}
\end{equation}
and the operator $K_{\mu}$
transforms as
\begin{equation}
K_{\mu} \rightarrow K_{\mu} 
+ i {g \over 16 \pi^2} \epsilon_{\mu \nu \alpha \beta}
\partial^{\nu} 
\biggl( U^{\dagger} \partial^{\alpha} U A^{\beta} \biggr)
+ {1 \over 96 \pi^2} \epsilon_{\mu \nu \alpha \beta}
\biggl[ 
(U^{\dagger} \partial^{\nu} U) 
(U^{\dagger} \partial^{\alpha} U)
(U^{\dagger} \partial^{\beta} U) 
\biggr].
\end{equation}
Gauge transformations shuffle a scale invariant operator 
quantity between the two operators $J_{\mu 5}^{\rm con}$ 
and $K_{\mu}$ whilst keeping $J_{\mu 5}^{GI}$ invariant.

The non-abelian three-gluon part of $K_+$ vanishes in $A_+=0$ gauge and 
the forward matrix elements of $K_+$ are invariant under residual gauge 
degrees of freedom in this gauge.
Furthermore
the forward matrix elements of $K_+$ measure the amount of spin carried
by gluonic partons in the target \cite{jaffe}.
This leads ultimately to the ``partons'', AB and JET 
scheme decompositions of $g_A^{(0)}$ in Section 2.
As soon as we go away from the forward direction matrix elements of
$K_+$ will pick up a gauge-dependent contribution, which must decouple
from any physical observable.

\subsection{The axial anomaly and higher moments}

The anomaly is also present in the $C=+1$ higher spin axial tensors
\cite{sb92}.
In general, for a given choice of renormalisation prescription $R$, 
the renormalised axial tensor operator differs from the gauge 
invariant operator by a multiple of a gauge-dependent, 
gluonic counterterm $K_{\mu \mu_1 ... \mu_{2n}}$, viz.
\begin{eqnarray}
\biggl[
{\overline q} \gamma_{\mu} \gamma_5 i D_{\mu_1} ... i D_{\mu_{2n}} q
\biggr]^R_{Q^2} (0) &=& \\ \nonumber
\biggl[
{\overline q} \gamma_{\mu} \gamma_5 i D_{\mu_1} ... i D_{\mu_{2n}} q
\biggr]^{GI}_{Q^2} (0)
&+& 
\lambda_{R, n}^{(K)} \ \biggl[ K_{\mu \mu_1 ... \mu_{2n} } \biggr]_{Q^2} (0)
+ 
\lambda_{R, n}^{(G)} \
\biggl[ G_{\mu \alpha} i D_{\mu_1} ... {\tilde G}^{\alpha}_{ \ \mu_{2n}} 
\biggr]^{(GI)}_{Q^2} (0)
\end{eqnarray}
Shifting between possible renormalisation schemes, we pick up contributions 
from counterterms involving the gauge-invariant gluonic operators
$G_{\mu \alpha} i D_{\mu_1} ... i D_{\mu_{2n-1}} 
 {\tilde G}^{\alpha}_{\mu_{2n}}$
and also the gauge-dependent $ K_{\mu \mu_1 ... \mu_{2n} }$.
The coefficients $\lambda_{R, n}^{(K)}$ and $\lambda_{R, n}^{(G)}$
are fixed by the choice of renormalisation prescription.
In $A_{+}=0$ gauge the two-gluon part of $K_{\mu \mu_1 ... \mu_{2n}}$ 
reads
\begin{equation}
K_{+...+(2n+1)} = 
{\alpha_s \over \pi} \lambda_{n}^{(K)} \
\epsilon_{+ \lambda \alpha \beta} \
A^{\alpha} \partial^{\lambda} (i \partial_{+})^{2n} A^{\beta }
\end{equation}
It is not easy to derive the $K_{\mu \mu_1 ... \mu_{2n}}$ beyond
the two gluon term.
Unlike the 
topological charge density 
${\alpha_s \over 4 \pi} G_{\mu \nu} {\tilde G}^{\mu \nu}$
the higher spin operators
${\alpha_s \over 4 \pi} 
G_{\mu \nu} i D_{\mu_{1}} ... i D_{\mu_{2n}} {\tilde G}^{\mu \nu}$ 
are not topological invariants for $n \geq 1$.
It follows that they are not total derivatives and, therefore, 
one
cannot use a divergence equation alone to fix the non-abelian 
part of $K_{\mu \mu_1 ... \mu_{2n}}$.
There is no equation
$\partial^{\mu_j} {\cal S} {\cal K}_{\mu \mu_1 ... \mu_{2n}}
 =
 {\alpha_s \over 4 \pi} 
 G_{\mu \nu} i D_{\mu_{1}} ... i D_{\mu_{2n}} {\tilde G}^{\mu \nu}$
for $n \geq 1$.

\section{The axial anomaly in ${\tilde F_q}$}

The first observation to make is that the non-forward matrix elements of 
$K_+$ (and $K_{+...+{\rm (2n+1)}}$) are gauge dependent in $A_+=0$ gauge, 
even in perturbation theory.
This means that these operators must decouple from any factorisation 
scheme or ``generalised operator product expansion'' \cite{jidvcs} 
for hard exclusive reactions like deeply virtual Compton scattering.
This is in contrast to the situation in deep inelastic 
scattering where the forward matrix elements of $K_+$
are invariant under residual gauge degrees of freedom in $A_+=0$ gauge.

Since there is no gauge-invariant local gluonic operator 
with quantum numbers $J^{PC} = 1^{++}$ it follows that 
the first moment of the spin-dependent {\it non-forward} 
gluonic coefficient must vanish.
This holds true in the non-forward generalisation of the 
${\overline {\rm MS}}$ factorisation scheme for handling 
the infra-red mass singularities.
\footnote{
See eg. the calculations of Mankiewicz et al. \cite{lech} 
and Ji and Osborne \cite{jidvcs};
in the notation of Mankiewicz et al. \cite{lech}
${\partial \over \partial u} C_1^{A,g} \bigg|_{u=0} = 0$.} 
However, there is no gauge-invariant non-forward generalisation of 
the popular JET and AB schemes used to describe polarised deep inelastic 
data.
In these schemes there is an explicit gluonic contribution 
to the first moment of $g_1$ associated with the invariant 
contribution of $K_+$ (in $A_+=0$ gauge) to the forward matrix 
element of $J_{+ 5}^{GI}$ induced by a non-vanishing first moment 
of the spin-dependent gluonic coefficient.

The gauge dependence of the non-forward matrix elements of $K_+$
also means that one has to be careful with the interpretation of
the spin-dependent non-forward parton distributions in terms of 
amplitudes to extract and then re-insert a (well defined) parton 
with a given momentum.
Consider 
the non-forward spin-dependent quark distribution defined first
using gauge invariant renormalisation and second via the partially
conserved axial vector current.
In the first case, gluonic information is intrinsically built into
the definition of the ``polarised quark'' via the axial anomaly.
In the second, the notion of ``polarised quark parton'' 
is no longer gauge invariant.
Taken together, this means that one has to be careful how far one 
carries 
through the semi-classical interpretation of parton distributions.
The gluonic information built into the flavour-singlet part of
${\tilde F}_q$ is manifest through the massive $\eta'$ and absence 
of any (nearly-)massless pseudo-Goldstone axial U(1) pole in ${\tilde E}_q$.

The zero-modes associated with dynamical $U_A(1)$ symmetry breaking 
mean that the perturbative QCD formula (5) for deeply virtual 
Compton scattering may not necessarily exhaust the total cross-section.
For example, in Crewther's theory of quark-instanton interactions one
would expect $\delta (x)$ and $\delta (x-\zeta)$ contributions in the
flavour-singlet part of ${\tilde H}_q(x,\zeta,t)$ 
associated the transfer of incident axial-charge carried by quarks and 
antiquarks respectively from partons with finite momentum to zero-modes. 
Such zero-mode contributions are not easily 
reconcilable with the perturbative QCD 
factorisation expression (5), in particular the singularities in the 
leading-twist coefficient function ${\tilde C}_q$ in Eq.(6).
This problem is not relevant to isovector pion or rho production 
\cite{piller}
which is described just in terms of the isovector non-forward 
distributions.
Any flavour-singlet zero-modes will not contribute to these processes.

\section{Orbital angular momentum}

A sum-rule \cite{jidvcs,shore} relates the form-factors appearing 
in $F_q$ and $F_g$ in the spin-independent part of 
${\cal A}_{\rm DVCS}$
to the quark and gluonic total angular-momentum contributions 
to the spin of the proton. 
The second moments of $F_q$ and $F_g$ project out the non-forward 
matrix elements of the QCD energy momentum tensor:
\begin{eqnarray}
& & \langle P' | \ T^{\mu \nu}_{q,g} \ | P \rangle  =
{\bar U}(P') \biggl[ 
  A_{q,g}(t) \gamma^{( \mu}{\bar P}^{\nu )} 
+ B_{q,g}(t) {\bar P}^{( \mu} i \sigma^{\nu ) \alpha} \Delta_{\alpha} / 2M
\\ \nonumber 
& & \ \ \ \ \ \ \ \ \ \ \ \ \ \ \ \ \ \ \ \ \ \ \ \ \ \ \ \ \ \ 
+ C_{q,g}(t) \Delta^{( \mu} \Delta^{\nu )} / M + {\cal O}(\Delta^3)
\biggr] U(P)
\end{eqnarray}
There are no massless bosons which couple to $T^{\mu \nu}$ in QCD,
which is associated with the fact that Poincare invariance is not 
spontaneously broken in QCD (with corresponding Goldstone bosons).
This means that the expansion in $\Delta$ and the forward limit of 
the form-factors in Eq.(28) are well-defined: 
there are no $\Delta_{\mu} \Delta_{\nu} / \Delta^2$ terms.
The current associated with Lorentz transformations is
\begin{equation}
M_{\mu \nu \lambda} = z_{\nu} T_{\mu \lambda} - z_{\lambda} T_{\mu \nu}
\end{equation}
The total angular momentum operator is related to the energy-momentum 
tensor by 
\begin{equation}
J_{q,g}^z = 
\langle P', {1 \over 2} | \
\int d^3z \ ( \vec{z} \ {\rm x} \ {\vec T}_{q,g} )^{z} \ 
| P, {1 \over 2} \rangle
\end{equation}
Substituting Eq.(28) into Eq.(30) and taking the forward limit
$\Delta \rightarrow 0$ 
one obtains
\begin{equation}
J^z_{(q,g)} = {1 \over 2} \biggl[ A_{q,g}(0) + B_{q,g}(0) \biggr]
\end{equation}
This result is Ji's sum-rule \cite{jidvcs} relating total angular 
momentum to the form-factors measured appearing in Eqs.(7) and (8).

If one can extract $J_{q,g}$ from the forward limit of $A(t)$ and $B(t)$ 
in hard exclusive processes, 
then subtracting
$S_z$
from polarised deep inelastic scattering 
would give information about the orbital angular momentum $L_z$.
Whereas the intrinsic spin $S_z^q$ is sensitive to the axial anomaly, 
the total angular momentum $J_z^{(q+g)}$ is not because it is measured
by a conserved current.
Theoretical studies \cite{ccm,orbital} show that 
$J_z^q$ and $J_z^g$ are each anomaly free in perturbative QCD 
meaning that the axial anomaly cancels between $S_z^q$ and $L_z^q$ 
in perturbation theory \cite{ccm,orbital,shore}.

This result generalises beyond perturbation theory \cite{shore} 
with the added consequence that there is no zero-mode 
contribution to $J_q$ and $J_g$. 
To see this, first consider the crossing symmetry in $x$ 
of the spin-independent quark distributions. 
The second moment of the charge parity plus distribution 
$(q + {\bar q})(x)$ 
projects out the nucleon matrix element of the energy-momentum tensor 
$T_{\mu \nu}$.
Any zero mode contribution to the right-hand side of 
the energy-momentum sum-rule would be associated 
with a $\delta'(x)$ term in $(q + {\bar q})(x)$.
Less singular $\delta (x)$ terms may be induced 
in the spin-dependent distributions by quark instanton interactions.
I know of no model which predicts a stronger singularity 
like $\delta '(x)$. 
Phenomenologically, such a term in the spin-independent 
structure function $F_2$ would lead to a violation of 
the energy-momentum sum-rule for partons, which is not observed.
Since there is no zero-mode contribution to $J_z$ it follows that any 
zero-mode contribution to $g_A^{(0)}$ is compensated by a second
zero-mode with equal magnitude but opposite sign in the orbital angular 
momentum $L_z^q$.
This has the practical consequence that one has to be careful how one 
interprets any determination of $L_z^q$ from DVCS through the sum-rule 
(31).
The orbital angular momentum carried by constituent quarks and 
by ``current quark'' partons are not necessarily the same, and 
is distinguished by the topological term ${\cal C}$ measurable 
in $\nu p$ elastic scattering.
What happens to spin and orbital angular momentum in 
the transition from current to constituent quarks is 
intimately related to the dynamics of axial U(1) symmetry breaking.

Finally, we note that in perturbative QCD one also has to be careful 
to quote any value of ``$L_z^q$'' with respect to the factorisation 
scheme used to extract ``$S_z^q$'' from polarised deep inelastic data.
For example,
$\Delta q_{\rm {\bar{MS}}} 
= 
( \Delta q  - {\alpha_s \over 2 \pi} \Delta g )_{\rm AB, \ JET}$
-- see also Shore \cite{shore00}.

\section{Conclusions}

In summary, the axial anomaly is manifest in deeply virtual Compton
scattering directly through the flavour-singlet spin-dependent part 
of the scattering amplitude ${\cal A}_{\rm DVCS}$ 
through the non-forward parton distribution 
$({\tilde F}_u + {\tilde F}_d +{\tilde F}_s)$ and indirectly through
the interpretation of information about quark 
orbital angular-momentum extracted from the spin-independent part of
${\cal A}_{\rm DVCS}$.

Whether QCD factorisation provides a complete description of the 
flavour-singlet, spin-dependent part of ${\cal A}_{\rm DVCS}$
depends on possible zero-mode contributions to
$({\tilde F}_u + {\tilde F}_d +{\tilde F}_s)$
which may be generated by the dynamics of axial U(1) symmetry
breaking.
In the absence of such zero-mode contributions perturbative QCD 
factorisation works for this term provided one chooses a 
gauge-invariant factorisation procedure such as minimal subtraction.
However
the resultant spin-dependent non-forward distribution will 
not have a simple canonical interpretation in terms of
extracting and re-inserting a quark or antiquark in the
target because gluonic information is intrinsic to 
$({\tilde F}_u + {\tilde F}_d +{\tilde F}_s)$ through the anomaly.
This gluonic information is manifest in the large-mass $\eta'$
pole in
$({\tilde E}_u + {\tilde E}_d +{\tilde E}_s)$
associated with the flavour-singlet pseudoscalar form-factor.

QCD anomaly effects cancel between the intrinsic spin and orbital
angular-momentum contributions to the total quark angular-momentum
in the proton.
This extends to possible zero-mode contributions associated with 
dynamical $U_A(1)$ symmetry breaking.
In QCD some fraction of the intrinsic spin carried by low-energy
constituent
quark quasi-particles may be carried by a zero-mode 
in addition to the partonic contributions from finite Bjorken $x$.
This zero-mode is compensated by a zero-mode with equal magnitude
and opposite sign in the orbital angular-momentum so that the total
quark angular-momentum is independent of the details of the current 
to constituent quark transition through ${\cal C}$
whereas the separate intrinsic and orbital contributions are not.
Finally, any value of the orbital angular-momentum extracted from
future DVCS experiments should also be quoted with respect to the
perturbative QCD factorisation scheme used to extract the partonic
intrinsic spin from polarised deep inelastic data.

\vspace{1.0cm}

{\bf Acknowledgements}

It is a pleasure to thank S.J. Brodsky, A. Donnachie, P.V. Landshoff, 
G. Piller, A. Radyushkin and W. Weise for helpful discussions.  
This work was supported in part by DFG.

\end{document}